\documentclass[preprint,showpacs,preprintnumbers,amsmath,amssymb,superscriptaddress]{revtex4}

\usepackage{graphicx,graphics}   

\usepackage{graphicx,amsfonts}
\usepackage{epsfig}

\usepackage{dcolumn}
\usepackage{bm}
\begin{document}

\title{Long lived hadronic resonances with exotic electric charge}

\author{G. De Conto}%
\email{georgedc@ift.unesp.br}
\affiliation{
Instituto  de F\'\i sica Te\'orica--Universidade Estadual Paulista \\
R. Dr. Bento Teobaldo Ferraz 271, Barra Funda\\ S\~ao Paulo - SP, 01140-070,
Brazil
}

\author{V. Pleitez}%
\email{vicente@ift.unesp.br}
\affiliation{
Instituto  de F\'\i sica Te\'orica--Universidade Estadual Paulista \\
R. Dr. Bento Teobaldo Ferraz 271, Barra Funda\\ S\~ao Paulo - SP, 01140-070,
Brazil
}

\date{09/23/2017}

\begin{abstract}
We classify the meson and baryon long lived resonances that may exist if quarks with electric charge 5/3 and $-4/3$ (in units of $\vert e\vert$) predicted by some 3-3-1 models do exist.  Some of these exotic resonances have the usual electric charges or $\pm(3,4,5)$, and the lightest ones decaying only into leptons plus known resonances. We propose another heavy $SU(3)_H$ global symmetry under which hadrons involving only exotic quarks can be constructed.
\end{abstract}


\maketitle

\section{Introduction}
\label{sec:intro}

In several instances, quesions that are still open in the Standard Model (SM) - ranging from neutrinos, collider physics, astrophysics and cosmology - are addressed by adding extensions to the SM, each one targeting a specific problem. This method however ends up producing a patchwork of possible solutions, lacking unity and consistency over distinct problems. The 3-3-1 models, which are based on the symmetry $SU(3)_C \times SU(3)_L \times U(1)_Y$, provide a unified framework under which several questions still open in the SM can be addressed and also bring new phenomena to be experimentally explored. This is the case of the minimal 3-3-1 model which contains only the known leptons but predicts new quarks, scalars and gauge bosons~\cite{Pisano:1991ee,Frampton:1992wt,Foot:1992rh}. Some of the features presented by this model are:
\begin{enumerate}
\item It explains the number of generations~\cite{Pisano:1991ee,Frampton:1992wt,Foot:1992rh} and the quantization of the electric charge~\cite{deSousaPires:1998jc},
\item It predicts singly and doubly charged bileptons, i.e. with lepton number $L=2$ \cite{DeConto:2015eia,Liu:1993gy},
\item It naturally incorporates the Peccei-Quinn symmetry~\cite{Pal:1994ba}, 
\item It predicts exotic particles, as quarks with exotic electric charge. Some consequences  of the latter ones are the subject of this work,
\item Incorporates many of the multi-Higgs extensions of the standard model singlets (neutral or charged), several doublets, and triplets~\cite{DeConto:2015eia}, 
\item It has many sources of $C\!P$ violation~\cite{Promberger:2007py}, 
\item It has a neutral $Z^\prime$ vector boson which contributes to flavor physics~\cite{Buras:2016egb},
\item The model has candidates for dark matter~\cite{Dong:2015rka},
\item The model here considered allows partial dynamical symmetry breaking in the vector bosons and known quarks~\cite{Das:1999hn}.
\end{enumerate}

A distinctive feature of these 3-3-1 models is that they predict the existence of new hadrons due to the exotic quarks present in the model. Some works have already discussed the experimental signatures of 3-3-1 models \cite{Alves:2012rb, Nepomuceno:2016jyr, Corcella:2017dns,Cao:2016uur}, mostly focusing on its bileptons, except for Ref. \cite{Cao:2016uur} which has a broader scope. However, we have not found any references that studied hadrons with the exotic quarks of the 3-3-1 models, nor hadrons with electric charge $\pm(3,4,5)$, which we find to be one of the most distinctive features of these models.

\section{New hadronic resonances in 3-3-1 models}
\label{sec:intro}

The minimal 3-3-1 model (m331)~\cite{Pisano:1991ee,Frampton:1992wt,Foot:1992rh} and the model with heavy charged leptons (331HL)\cite{Pleitez:1992xh} have quarks carrying non-standard values of the electric charge (in units of $\vert e\vert$).  One with electric charge $5/3$, denoted by $J$,  and two with electric charge $-4/3$, denoted by $j_{1,2}$. These quarks are probably heavy and as in the case of $D$ and $B$ mesons and baryons any flavor symmetry larger than $SU(3)$ is badly broken. However, as in the case of the $c$ and $b$ quarks we use the flavor symmetry in the light quark flavors $u,d,s$ to classify the new mesons and baryons as $SU(3)$ multiplets. 

Let us consider first hadrons with only one exotic quark. The new mesons and baryons may be of the form $Q\bar{q}$ and $Qq_1q_2$, where $Q$ denotes $J$ or $j_{1,2}$ and $q_{1,2}$ denote an usual quark. For instance, the possible $j$-mesons are in anti-triplet $\textbf{3}^*$  of the global $SU(3)$ symmetry of the light flavor $u,d,s$ where we replace one of the known quarks with an exotic quark, see Table~\ref{table1}.  Baryons are in $\textbf{3}^*$ and $\textbf{6}$ of $SU(3)$ shown in Table~\ref{table2} and Table~\ref{table3}, respectively. Also there are the following triplets involving two exotic quarks: one triplet $(JJu)^{+4},(JJd)^{+3},(JJs)^{+3}$, two triplets $(Jj_au)^{+1}(Jj_ad)^0,(Jj_as)^0$, and three triplets $(j_aj_bu)^{-2},(j_aj_bd)^{-3},(j_aj_bs)^{-3}$, where $a,b=1,2$ (see Table \ref{table4}). The superscript denotes the electric charge of the hadron.
We also have hadrons involving only exotic quarks: mesons $(\bar{J}J)^0, (\bar{J}j)^{-3},(\bar{j}j)^0$ and baryons  $(JJJ)^{+5},(JJj)^{+2},(Jjj)^{-1},(jjj)^{-4}$, where $j$ may be $j_1$ or $j_2$ (sse tables \ref{table5} and \ref{table6}). Notice that there are resonances with electric charge $\pm(3,4,5)$ which do not exist in the SM and many of its extensions.

The heavy hadrons involving only the exotic quarks can be classified by a global heavy  $SU(3)_{H}$ symmetry  introducing a heavy strangeness $S_H$, and a Heavy hypercharge, $Y_H=B+S_H$, where $B$ is the baryon number. In this case the Gell-Mann-Nishijima formula is $Q_H=\kappa T_{3H}+Y_H/2$. The $I_H$ doublet is $(j_1,J)$, with $S_H=0, Y_H=B=1/3$,  and $\kappa=-3$; and the isosinglet, $j_2$, has $S_H=-3$. For instance the nonet of mesons are composed by a triplet $\Pi^0(\bar{J}J,\bar{j}_1j_1),\Pi^{-3}(\bar{J}j_1),\Pi^{+3}(\bar{j}_1J)$, a doublet $\mathcal{K}^{-3}(\bar{J}j_2),\mathcal{K}^0(\bar{j}_1j_1)$ and the respective anti-doublet, the singlet with $S_H=0$, $\eta_J(\bar{J}J,\bar{j}_1j_1,\bar{j}_2j_2)$, and finally  the singlet with $S_H=-6$, $\eta^\prime_J(\bar{J}J,\bar{j_1}j_1,\bar{j}_2j_2)$. 
These degrees of freedom may be excited at energies above the Landau pole existing in these 3-3-1 models at a few TeVs~\cite{Dias:2004dc}. 

\begin{center}
\begin{figure}
\centering
\includegraphics[scale=0.3]{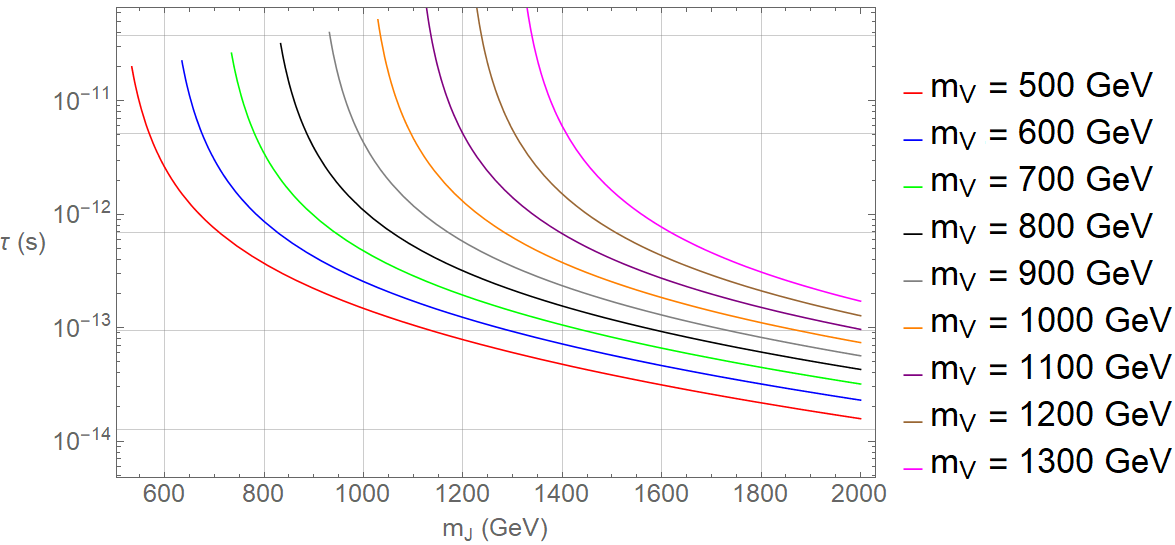}
\caption{Exotic quark lifetime assuming a decay into an ordinary quark (up quark in this case) and a $V$ vector boson. See Eq.~(\ref{largura1}). In this plot the horizontal axis indicates the exotic quark mass and each line corresponds to  different $V$ masses.}
\label{fig:resultF1}
\end{figure}
\end{center}

All these exotic hadrons may be long lived  because they decay only through leptonic decays. Just to illustrate typical values of the half-life of these new hadrons, here we will not consider all the contributions that can affect this half-life. For instance, at lowest order, the decay width for an exotic quark decaying exclusively into an ordinary quark and a $V$ vector boson ($V$ may be singly or doubly charged) is given by:
\begin{eqnarray}
\Gamma(J\to Vq)&=& g\,K_V\frac{\vert F_{Jq}\vert^2}{8\pi m_J\hbar c}\left[\frac{2E_V}{M^2_V}(E_qE_V+K_V^2)\right.\nonumber \\ &&+\left. E_q\left( 2-\frac{E^2_V-K_V^2}{M^2_V} \right)\right]\nonumber\\&&
\label{largura1}
\end{eqnarray}
where $4m^2_J\,K_V^2=[(m_J-(m_q+M_V)^2)][(m_J-(m_q-M_V)^2)]$, $g$ is the electromagnetic coupling constant ($g=\sqrt{4 \pi \alpha}/s_W$), $E_V$ and $E_q$ are the vector boson and quark energies ($E_i=\sqrt{m_i^2+K_V^2}$). If $V=V^+$then $q=u,c,t$; and if $V=U^{++}$ then $q=d,s,b$. $F_{Jq}$ is a dimensionless parameter related to the coupling of the $J$ and $q$ quarks, which we assume, just to illustrate, to be $F_{Jq}=0.1$ so that perturbative calculations are possible.  Using $m_q=m_u=2.3$ MeV, we obtain the lifetimes $\tau_J$ which are shown in Fig.~\ref{fig:resultF1}.  Notice that they may be of the order of $10^{-14}$ s. 

\begin{center}
\begin{figure}
\centering
\includegraphics[scale=0.3]{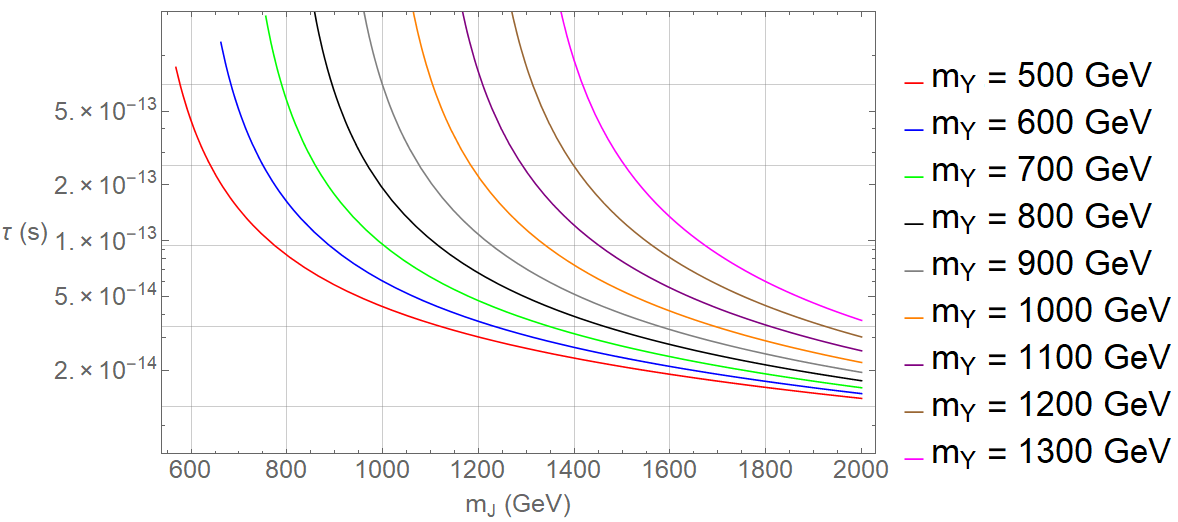}
\caption{Same as in Fig.~\ref{fig:resultF1} but considering a scalar $Y$ and  using Eq.~(\ref{largura2}). 
}
		\label{fig:resultF2}
	\end{figure}
\end{center}

There are also contributions of singly and doubly charged scalars denoted by $Y$.
We can consider also generic scalar and pseudo-scalar interactions with fermion-scalar vertices written as $i(S_Y+P_Y\gamma_5)$. In this case, once again at lowest order, assuming that the exotic quark decays only into scalar-quark pairs
\begin{eqnarray}
\Gamma(J \to Yq)&=& \frac{K_Y}{8 \pi  m_J^2 \hbar c} \left[ 4 m_J E_q (|S_Y|^2+|P_Y|^2)\right.\nonumber \\ &&+\left. m_J m_q (|S_Y|^2-|P_Y|^2) \right]\nonumber\\&&
\label{largura2}
\end{eqnarray}
where $4m^2_J\,K_Y=[(m_J-(m_q+M_Y)^2)][(m_J-(m_q-M_Y)^2)]$. Assuming $S_Y=P_Y=0.1$, once again so that perturbative calculations are possible, we obtain the lifetimes shown in Fig.~\ref{fig:resultF2}.

The mesons $\bar{J}J$ would be a non-relativistic bound state and assuming that it is in a Coulombian potential,
the ratio of the period of $\bar{J}J$ with respect to the $\bar{t}t$ is $T_{\bar{J}J}/T_{\bar{t}t}=m_t/m_J$, Even if $M_J=m_t$, $T_{\bar{J}J}\sim10^{-25}$ s~ (the $\bar{t}t$ lifetime discussion can be found in \cite{Donoghue:1992dd}). Hence, before $J$ or $\bar{J}$ decays there is enough time to form a ``$J$-tonium" bound state.  

In the case of the scalar or vector fields being heavier than the $J,j$ quarks, there are Fermi-like effective interactions, for instance $G_{HF}\bar{J}Oq\bar{l}O^\prime \nu$, with $O,O^\prime$ including $\gamma^\mu\gamma_5$ or $\gamma_5$ depending if the heavy mediator is a vector or a scalar.

Neutral resonances as $\Pi^0(\bar{J}J,\bar{j}j)$ may be produced mainly through gluon fusion and they can decay into two photons:
\begin{equation}
\Gamma_{\Pi^0\to \gamma\gamma}=N_c\frac{\alpha}{192\pi^2}\frac{m^3_{\Pi^0}}{v_\chi^2},
\label{Pion}
\end{equation}
where $v_\chi$ is the vacuum expectation value that gives mass to the exotic particles in the m331. With $N_c=3$ and assuming $m_{\Pi^0}\approx v_\chi=1$ TeV we have $\tau_\Pi=5.7 \times 10^{-23}$ s.

\begin{center}
	\begin{figure}
		\centering
		\includegraphics[scale=0.3]{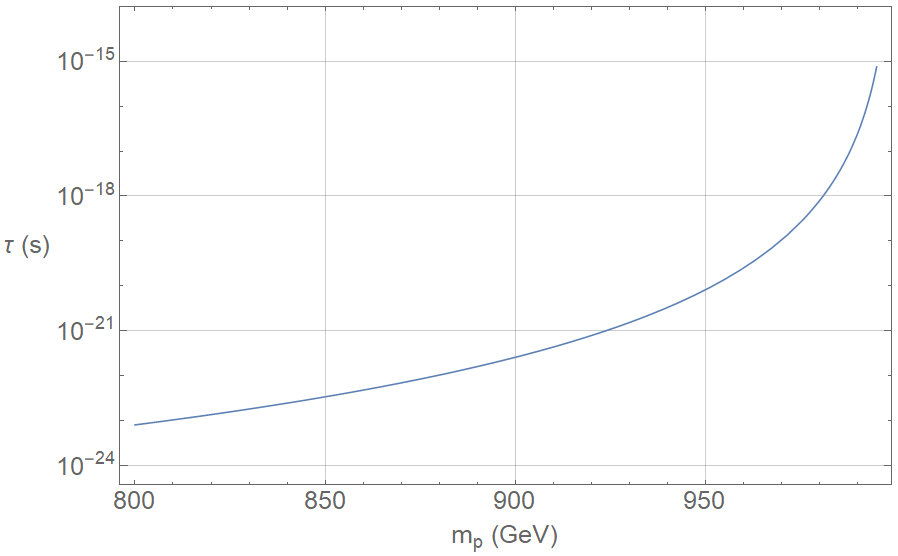}
		\caption{The lifetime of the beta-like decay between two exotic resonances. Here $m_p$ denotes the mass of a proton-like resonance.
		}
		\label{fig:betadecay}
	\end{figure}
\end{center}

It is possible that one of the baryon resonances involving only exotic quarks decays into a lighter one with the emission of a lepton and a neutrino. Considering a decay similar to the neutron beta decay, for instance $(Jjd)^0\to (Jdd)^++e^-+\bar{\nu}_e$, and using the usual expression for the neutron case with $M_V=M_W$ and $V_{ud}=K$, where $K$ is the mixing matrix in the vertex $jd$, we obtain a lifetime of the order $(10^{-15}-10^{-23})$ s depending on the mass difference between the resonances. See Fig.~\ref{fig:betadecay}. This is indeed a lower limit of the lifetime since $M_V\gg M_W$ and $K\not= V_{ud}$. There are other decays such as $\Pi^0(\bar{J}J)\to \gamma(Z,Z^\prime)\to\bar{f}f$, where $f=\nu,l,u,d,\cdots$ where the ellipses denote quarks lighter than $m_{\Pi^0}/2$. Recall that the interactions with $Z^\prime$ are flavor changing.

Finally, we stress that these exotic quarks are almost stable thanks to an automatic $\mathbb{Z}_2$ symmetry in the quark-scalar-vector interactions. This symmetry is broken in the lepton interactions in the minimal 3-3-1 model (there exist the interactions $V-l-\nu$ and $U-l-l$) and in the scalar sector through a quartic term $\chi^\dagger\eta\rho ^\dagger\eta$, where $\eta,\rho,\chi$ are scalar triplets needed in both the m331 model~\cite{Pisano:1991ee,Frampton:1992wt,Foot:1992rh} and the 331HL \cite{Pleitez:1992xh}. In the 3-3-1 model with heavy charged leptons~\cite{Pleitez:1992xh} the $\mathbb{Z}_2$ symmetry is explicitly broken only in the quartic scalar interaction mentioned above. 
In the former model there are also trilinear and quartic terms involving the scalar sextet that are needed to give the correct masses in the lepton sector with, or without, sterile neutrinos~\cite{DeConto:2015eia}. 
The $\mathbb{Z}_2$ symmetry implies that the resonances involving at least one exotic quark $J$ or $j$ are long lived and the lightest ones, independently of the electric charge, would only decay producing leptons plus ordinary hadrons. For instance, in the m331 model we have:  $J\to V^+ u\to l^+\nu u$, or $J\to U^{++}d\to l^+l^+d$; and $j\to V^-d\to l^-\nu_ld$ or $j\to U^{--}u\to l^-l^-u$. 

\section{Conclusions}
\label{sec:con}

We have shown that in the 3-3-1 models that we have considered there are new hadrons which are almost stable, and some of them having exotic electric charge $\pm(3,4,5)$.  In fact, as we said before, the possible existence of these resonances is a distinctive feature of the minimal 3-3-1 model and the 3-3-1 model with heavy charged leptons, since in almost all proposed extensions to the SM they do not exist. However, indirect effects of these new hadrons can be observed in lepton-nucleon scattering. For instance, in neutrino-nucleus interactions the process $\nu_L+N\to \mu^+_L+X$ ($X$ is anything) fakes an anti-neutrino interaction if the helicity of the charged muon is not measured.

It is important to notice the following. 1) Since these resonances could have existed in the early universe, eventually they would decay into leptons plus usual hadrons, and having the 3-3-1 models several sources of $C\!P$ violation~\cite{Promberger:2007py}, those decays may induce an asymmetry  between leptons and anti-leptons with or without heavy neutrino decays. For instance $\Gamma(J\to l^+\nu u)$ may be different from $\Gamma(\bar{J}\to l^-\bar{\nu} \bar{u})$, and $J\to l^+l^+d$ different from $\bar{J}\to l^-l^-\bar{d}$. 2) The exotic neutral vector boson in these models is leptophobic (the $Z^\prime-l-l$ couplings are proportional to $\sqrt{1-4s^2_W}$ \cite{Dias:2006ns}) and in this case the decays of $Z^\prime$ are expected to be  $Z^\prime\to f\bar{f}$~\cite{Rosner:1996eb} where $f$ is a known heavy quark ($t$) or a heavy exotic quark ($J,j$) or a heavy lepton $E^+$ in the case of the 331HL model.\\


\noindent\textit{Acknowledgments.---} The authors would like to thanks to Conselho Nacional de de Desenvolvimento Cient\'{\i}fico e Tecnol\'{o}gico (CNPq) for full (G.D.C.) and partial (V.P.) support.  (V.P.) is also  thankful for the support of Funda\c{c}\~{a}o de amparo \`{a} pesquisa do estado de S\~{a}o Paulo (FAPESP) funding Grant No. 2014/19164-6. 


\newpage

\begin{table}[ht]
\begin{tabular}{|l|l|c|l|l|c|} \hline  
 Quarks $J$ &  Mesons-J & $S$  &Quarks $j$ & Mesons-j  & $S$ \\ \hline
  $J\bar{d}$  & $\mathcal{J}^{++}_{d}$ & 0 & $j\bar{d}$ & $j^{-}_d$ & $0$ \\ \hline
 $J\bar{u}$ &  $\mathcal{J}^{+}_u$ & 0 & $j\bar{u}$&$j^{--}_u$ &$0$  \\ \hline
 $J\bar{s}$ &  $\mathcal{J}^{+}_s$ & $+1$ &  $j\bar{s}$ & $j^{-}_s$ & $+1$  \\ \hline
\end{tabular}
\caption{Mesons in the representation $\textbf{3}^*$,~$S$~is the strangeness and $j$ may be~$j_1$ or $j_2$. The mesons-J have $S_H=0$ and the meson-j have $S_H=-3$.}
\label{table1}
\end{table}

\begin{table}[ht]
\begin{tabular}{|l|l|c|l|l|c|} \hline  
 Quarks $J$ &  Baryons-J & $S$ &Quarks $j$ & Baryon-j  & $S$  \\ \hline
  $Jud$  & $\Theta^{++}_{Jud}$ & $0$  & $jud$ & $\theta^{-}_{jud}$ & $0$\\ \hline
$Jus$ &  $\Theta^{+}_{Jus}$ &$-1$ & $jus$& $\theta^{-}_{jus}$ &$-1$ \\ \hline
 $Jds$ &  $\Theta^{+}_{Jds}$ & $-1$ & $jds$ & $\theta^{-}_{jds}$ & $-1$\\ \hline
\end{tabular}
\caption{Baryons in representation $\textbf{3}^*$. The $S_H$ asignment is as in Table \ref{table1}. 
}
\label{table2}
\end{table}

\begin{table}[ht]
	\begin{tabular}{|l|l|c|l|l|c|} \hline  
 Quarks $J$ &  Baryons-J & $S$ &Quarks $j$ & Baryon-j  & $S$ \\ \hline
  $Juu$  & $\Theta^{\prime +++}_{Juu}$ & $0$  & $juu$ & $\theta^0_{juu}$ & $0$\\ \hline
$Jud$ &  $\Theta^{\prime++}_{Jud}$ &$0$ & $jud$& $\theta^-_{jud}$ &$0$ \\ \hline
$Jdd$ &  $\Theta^{\prime +}_{Jdd}$ & $0$ & $jdd$ & $\theta^{--}_{jdd}$ & $0$\\ \hline
$Jus$ &  $\Theta^{\prime++}_{Jus}$ & $-1$ & $jus$  &  $\theta^-_{jus}$ & $-1$ \\ \hline 
$Jds$ &$\Theta^{\prime +}_{Jds}$ & $-1$  & $jud$ &  $\theta^{--}_{jds}$ & $-1$ \\ \hline 
$Jss$ &$\Theta^{\prime +}_{Jss}$ & $-2$  & $jss$ & $\theta^{--}_{jss}$  & $-2$\\ \hline 
	\end{tabular}
	\caption{Baryons in the representation $\textbf{6}$. The notation and the $S_H$ assignment are as in Table \ref{table2}. 
	}
	\label{table3}
\end{table}

\begin{table}
	\begin{tabular}{|l|c|l|c|l|c|} \hline
		Quarks $J$ & $S$ & Quarks $J$ & $S$ & Quarks $J$ & $S$ \\ \hline
		$JJu$ & 0 & $Jj_au$ & 0 & $j_a j_b u$ & 0 \\ \hline
		$JJd$ & 0 & $Jj_ad$ & 0 & $j_a j_b d$ & 0 \\ \hline
		$JJs$ & -1 & $J j_a s$ & -1 & $j_a j_b s$ & -1 \\ \hline
	\end{tabular}
\caption{Baryons with two exotic quarks, where $a,b=1,2$. The notation and the $S_H$ assignment are as in Table \ref{table1}.}
\label{table4}
\end{table}

\begin{table}
	\begin{tabular}{|l|c|l|c|} \hline
		Quarks $J$ & $S_H$ & Quarks $J$ & $S_H$ \\ \hline
		$JJJ$ & 0 & $JJj_a$ & -3 \\ \hline
		$J j_a j_b$ & -6 & $j_a j_b j_c$ & -9 \\ \hline
	\end{tabular}
	\caption{Baryons with three exotic quarks, where $a,b,c=1,2$. Notice that in this table we have $S_H$ instead of $S$.}
	\label{table5}
\end{table}

\begin{table}
	\begin{tabular}{|l|c|l|c|l|c|} \hline
		Quarks $J$ & $S_H$ & Quarks $J$ & $S_H$ & Quarks $J$ & $S_H$ \\ \hline
		$\bar{J} J$ & 0 & $\bar{J} j_a$ & -3 &	$\bar{j}_a j_b$ & 0 \\ \hline
	\end{tabular}
	\caption{Mesons with two exotic quarks, where $a,b=1,2$. Notice that in this table we have $S_H$ instead of $S$.}
	\label{table6}
\end{table}

\newpage

\newpage

\end{document}